\documentclass[twocolumn,preprintnumbers,amsmath,amssymb,superscriptaddress]{revtex4}
\usepackage{graphicx}
\usepackage{dcolumn}
\usepackage{bm}
\usepackage{soul}
\usepackage{color}
\usepackage{epstopdf}
\usepackage[version=3]{mhchem}
\usepackage{lipsum}
\usepackage[outercaption]{sidecap}
\usepackage{floatrow}
\begin{document}
\title{Confinement effects on the spatially inhomogeneous dynamics in  metallic glass films}% 
\author{Anh D. Phan}
\affiliation{Faculty of Materials Science and Engineering, Phenikaa Institute for Advanced Study, Phenikaa University, Hanoi 12116, Vietnam}
\email{anh.phanduc@phenikaa-uni.edu.vn}
%\author{Nguyen K. Ngan}
%\affiliation{Faculty of Materials Science and Engineering, Phenikaa University, Hanoi 12116, Vietnam}
\date{\today}

\date{\today}

\begin{abstract}
We develop the Elastically Collective Nonlinear Langevin Equation theory to investigate, for the first time, glassy dynamics in capped metallic glass thin films. Finite-size effects on the spatial gradient of structural relaxation time and glass transition temperature ($T_g$) are calculated at different temperatures and vitrification criteria. Molecular dynamics is significantly slowed down near rough solid surfaces and the dynamics at location far from the interfaces is sped up. In thick films, the mobility gradient normalized by the bulk value well obeys the double-exponential form since interference effects between two surfaces are weak. Reducing the film thickness induces a strong dynamic coupling between two surfaces and flattens the relaxation gradient. The normalized gradient of the glass transition temperature is independent of vitrification timescale criterion and can be fitted by a superposition function as the films are not ultra-thin. The local fragility is found to remain unchanged with location. This finding suggests that one can use Angell plots of bulk relaxation time and the $T_g$ spatial gradient to characterize glassy dynamics in metallic glass films. Our computational results agree well with experimental data and simulation. 
\end{abstract}

\maketitle
\section{Introduction}
After several decades of continuous debate, the activated dynamics of thin glass-forming films remains poorly understood due to complexities of  confinement effects and diverse applications \cite{3,4,8,14,20,21,22,23,24,17,15}. Vapor interfaces universally accelerates relaxation nucleated from free surfaces towards the interior. For solid interfaces, the spatially inhomogeneous dynamics becomes more complex and has less universal behaviors because surface roughness, film-substrate adsorption/attractions, and mechanical properties of substrate can significantly affect. The presence of solid surfaces can increase, decrease, or even do not modify bulk dynamics. These variations of dynamics strongly depend on specific details in design. Thus, it is essential to construct a predictive theoretical approach to understand the dependence of temperature and film thickness on the mobility gradients varies from the interface and transfers into the film.

For many years, Mirigian, Phan, and Schweizer has formulated different versions of the Elastically Collective Nonlinear Langevin Equation (ECNLE) theory \cite{2,7,10,6,35,42,11,61,62,44,45} to study quantitatively and qualitatively glassy dynamics of bulk amorphous materials including polymers, drugs, thermal liquids, metallic glasses, and thermal liquids at ambient and elevated pressures. An activated event in the ECNLE theory is governed by local motions within cages and long-range collective molecular dynamics surrounding the cage. Effects of the cage-scale dynamics on the relaxation process dominate those of the collective dynamics at high temperatures but this trend is reversed at low temperatures. The temperature dependence of structural relaxation time ranging 1 ps to $10^3$ s and dynamic fragility are calculated. The prediction enable to quantitatively and qualitatively understand physical phenomena in simulation (the timescale $\leq 10^6$ ps) and experimental observation (the timescale $\sim 1-100$ s).  

We extended the ECNLE theory to predict the spatial gradient of the alpha relaxation time and glass transition temperature in polymer film systems having vapor and solid surface \cite{3,4,8,24}. Our new treatment describes the spatial propagation of caging constraints from the surface into the film interior. The dynamical caging constraints can be significantly weakened, slightly softened, and hardened at the vapor, smooth solid, and rough solid interface, respectively. From these, we determine how mobility changes and explain why the relaxation time gradient for free-standing and supported polymer films obeys a "double-exponential" form suggested by simulations few decades ago \cite{17,15}. While finite-size effects on the inhomogeneous glassy dynamics in free-standing polymer films have been recently investigated \cite{24}, problems for capped films have not been addressed yet, particularly when material is metallic glass. It is still unclear how two solid interfaces cause nonadditive interference effects to modify mobility gradients and invalidate the "double-exponential" form, how strong the interfacial coupling is when the films are thinner, how good an ansatz superposition approximation \cite{17} is when modelling a normalized alpha relaxation time gradient of metallic glass films at different thicknesses.

In this paper, we develop the ECNLE theory to determine, for the first time, glassy behaviors of metallic glass thin film capped by two solid surfaces. The hard interfaces interact with the film via repulsive forces and, thus, can be treated as neutral rough surfaces. We construct a unified physical interpretation for influences of hard interface modified caging constraints on the gradients of mobility, glass transition temperature, dynamic fragility, and how interfacial effects from two surfaces are dynamically coupled when varying film thickness. Analytical expressions are clearly derived to explain numerical results. We also compare our findings with prior simulation and experiments.  

\section{Theoretical background}
According to the ECNLE theory \cite{2,7,10,6,35,42,11,61,62,44,45}, glass-forming liquids are modeled as a hard-sphere fluid. The dynamic free energy of a tagged particle in a bulk fluid is

\begin{eqnarray}
\frac{F_{dyn}^{bulk}(r)}{k_BT} &=&-3\ln\frac{r}{d}
\nonumber\\ &-&\int_{0}^{\infty} dq\frac{ q^2d^3 \left[S(q)-1\right]^2}{12\pi\Phi\left[1+S(q)\right]}\exp\left[-\frac{q^2r^2(S(q)+1)}{6S(q)}\right]\nonumber\\
&=& F_{ideal}(r) + F_{caging}^{bulk}(r),
\label{eq:1}
\end{eqnarray}
where $\Phi$ is the volume fraction, $k_B$ is Boltzmann constant, $T$ is temperature, $d$ is the particle diameter, $r$ is the displacement from the initial position, and $S(q)$ is the static structure factor of hard-sphere fluids calculated using the Percus-Yevick theory \cite{1}.

\begin{figure}[htp]
\includegraphics[width=8.5cm]{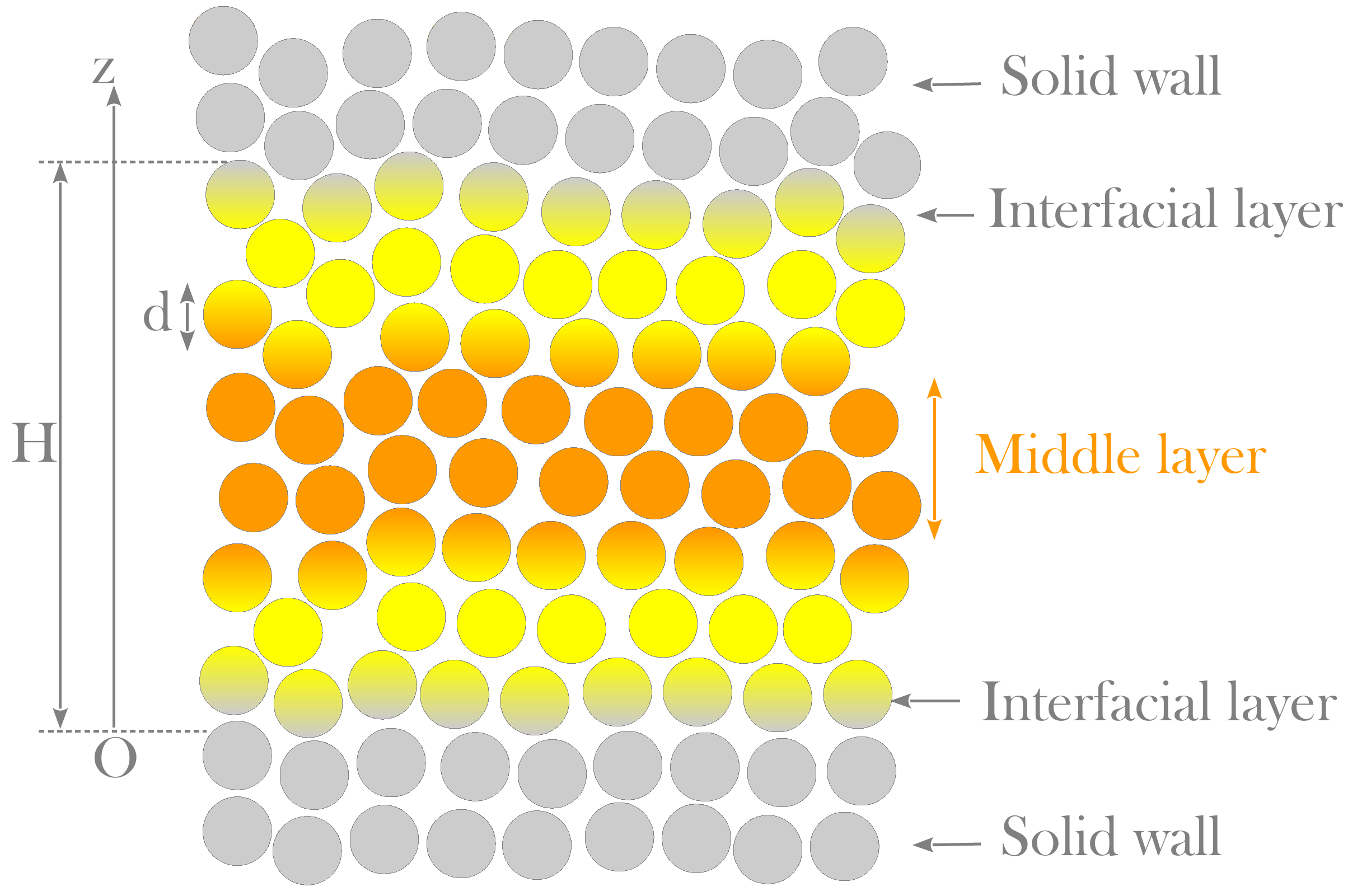}
\caption{(Color online) Schematic illustration of capped metallic films indicating the coordinate $z$ and film thickness $H$. Color of particles illustrates the mobility gradient of the thin polymer film with neutrally solid substrates.}
\label{fig:0}
\end{figure}

When investigating mobility gradients of one-interface thick films \cite{3,4,8}, the distance $(z)$ dependence of the dynamic free energy, $F_{dyn}(r,z)$, in the ECNLE theory is constructed using several assumptions. First, interfacial effects on packing structure (both pair correlation function and density profile) is ignored. This assumption is consistent with the definition of neutral confinement performed in simulation studies \cite{3,17,15,25,26}. This idea removes changes of structure as a complicating factor in trying to understand dynamical gradients. Thus, we can focus on a purely dynamic scenario with no perturbation in thermodynamics or the structure in the film. Second, a particle at the interface loses liquid nearest neighbors in its particle cage and the interface is very sharp. Missing particles due to solid surface are theoretically replaced by pinned hard-sphere particles, which have the size and density identical to that in the film \cite{12} as illustrated in Fig. \ref{fig:0}. In basic physics-oriented simulations \cite{17,15,25,26}, the interface is known as a type of "micro" roughness model. One can simulate a 1-component liquid, insert a dividing plane, and pin particles below it in their equilibrium configurations to create a pinned particle rough interface. In fact, this is a “micro-rough” interface since the surface is composed of precisely the same particles (with the same pair structure) as the mobile particles of the liquid. The presence of pinned particles solidifies caging constraints in the surface layer and effects of the surface caging force is spatially transferred into the film. Thus, the dynamical gradient occurs and is illustrated by different colors of particles. Here, we use the ECNLE theory for bulk fluids of pinned particles to calculate the surface caging potential under neutral confinement conditions \cite{12}. Third, in the film, the caging constraints acting on particles in layer $i$ is affected by the constraints in layer $i-1$. The distance from the interface is $z=(i-1)d$. This developed the ECNLE theory for thick films has provided good quantitative descriptions for simulations and experiments of polymer thick films with a vapor and rough solid interface \cite{3,4,8}. The dynamic free energy in $n^{th}$ of the thick film is
\begin{eqnarray}
F_{dyn}^{(n)}(r) &=& F_{ideal}(r) + \left(1-\frac{1}{2^n}\right)F_{caging}^{bulk}(r)  +
 \frac{F_{caging}^{surface}(r)}{2^n}, \nonumber\\
 &=& F_{dyn}^{bulk}(r) + \frac{F_{caging}^{surface}(r)-F_{caging}^{bulk}(r)}{2^n},\nonumber\\
  &=& F_{dyn}^{bulk}(r) + \frac{1}{2}\frac{\Delta F_{caging}(r)}{2^{z/d}},
\label{eq:2}
\end{eqnarray}
where $\Delta F_{caging}(r)=F_{caging}^{surface}(r)-F_{caging}^{bulk}(r)$ is the difference between the solid-surface-induced caging constraint and its bulk counterpart at the interface, and
\begin{widetext}
\begin{eqnarray}
F_{caging}^{surface}(r) &=& -2\int\frac{d\mathbf{q}}{(2\pi)^3}\left[ \frac{C(q)S_{12}(q)e^{-q^2r^2/6}}{\rho(1-\alpha)\left[1- \rho(1-\alpha)C(q)\right]}+
\frac{\rho(1-\alpha)C(q)^2e^{-q^2r^2\left[2-\rho(1-\alpha)C(q) \right]/6}}{\left[1- \rho(1-\alpha)C(q)\right]\left[2- \rho(1-\alpha)C(q)\right]} \right] \nonumber\\
&-& F_{caging}^{bulk}(r),
\label{eq:3}
\end{eqnarray}
\end{widetext}
where $\alpha$ is the fraction of pinned particles in the cage and $S_{12}(q)$ is the collective static structure factor between pinned and mobile particles. The analytical expression of $S_{12}(q)$ can be found in Refs. \cite{3,4,12}. Since particles at the interface lose one half of the nearest neighbors, $\alpha = 0.5$.

By using the same analysis, when the film capped by two solid surfaces has a finite-size thickness of $H$, the dynamic free energy becomes
\begin{eqnarray}
F_{dyn}(r,z,H) &=& \frac{\Delta F_{caging}(r)}{2}\left(\frac{1}{2^{z/d}}+\frac{1}{2^{(H-z)/d}}\right)\nonumber\\
&+& F_{dyn}^{bulk}(r).
\label{eq:4}
\end{eqnarray}
This dynamic free energy gives the local barrier, $F_B(z,H)= F_{dyn}(r_B,z,H)-F_{dyn}(r_L,z,H)$, the localization length or the minimum position of $F_{dyn}(r,z,H)$, $r_L(z,H)$, and the barrier position, $r_B(z,H)$, corresponding to the local maximum of $F_{dyn}(r,z,H)$. For free-standing polymer films, $F_{caging}^{surface}(r)=0$ since there is no particle in vapor layers, Eq. (\ref{eq:4}) recovers to the formula in Ref. \cite{24}, which is
\begin{eqnarray}
F_{dyn}(r,z,H) &=& \frac{ F_{caging}^{bulk}(r)}{2}\left(2-\frac{1}{2^{z/d}}-\frac{1}{2^{(H-z)/d}}\right)\nonumber\\
&+& F_{ideal}(r).
\label{eq:4-1}
\end{eqnarray}

In previous works \cite{45,9}, authors found that the collective elastic barrier characterizing for effects of cooperative motions on the glass transition is almost zero in bulk metallic glasses. We suppose the conclusion remains unchanged in the metallic glass films. Thus, the structural relaxation time is calculated by
\begin{eqnarray}
\frac{\tau_\alpha}{\tau_s} = 1+ \frac{2\pi}{\sqrt{K_0(z,H)K_B(z,H)}}\frac{k_BT}{d^2}\exp\left(\frac{F_B(z,H)}{k_BT} \right),
\label{eq:5}
\end{eqnarray}
where $K_0(z)$ and $K_B(z)$ are the absolute curvature of $F_{dyn}(r_B,z,H)$ at localization length and barrier position, respectively, and $\tau_s$ is short relaxation time analytically reported elsewhere \cite{2,7,10,6,35,42,11,61,62,44,45}.

The above calculations give us $\tau_\alpha(\Phi, z, H)$. To determine the temperature dependence of $\tau_\alpha$, we use a density-to-temperature conversion or thermal mapping \cite{45,9}, $T=T_g+(\Phi_g-\Phi)/\beta\Phi_0$, here $T_g$ is the glass transition temperature, $\Phi_g$ is the volume fraction corresponding to the vitrification timescale criterion of $T_g$ ($\tau_\alpha(\Phi_g)=\tau_\alpha(T_g)$), $\Phi_0=0.5$ is the characteristic volume fraction, and $\beta=12\times10^{-4}$ $K^{-1}$ is the effective thermal expansion coefficient. 

It is important to note that this is a minimalist approach to predict the temperature dependence of structural relaxation time and gradients of the glass transition temperature in finite-size films. Thus, we employ the hard-sphere fluid to describe dynamics of glass formers. Information of atomic interactions are encoded in the thermal mapping constructed using the thermal expansion. There is no adjustable parameter needed in our calculations. One can consider effects of intermolecular forces on structure and dynamics of amorphous materials by using the standard reference interaction site model (RISM) \cite{1} or other molecular dynamics simulations to obtain $S(q)$ and $g(r)$ for Eq. (\ref{eq:1}-\ref{eq:4-1}) and calculate $\tau_\alpha(T,z)$. However, we have to know a mathematical form and its parameters of the intermolecular potential, unless ECNLE calculations have several control parameters and the thermal mapping is also altered. The problem becomes very complicated.

\section{Results and Discussion}
\begin{figure}[htp]
\includegraphics[width=8.5cm]{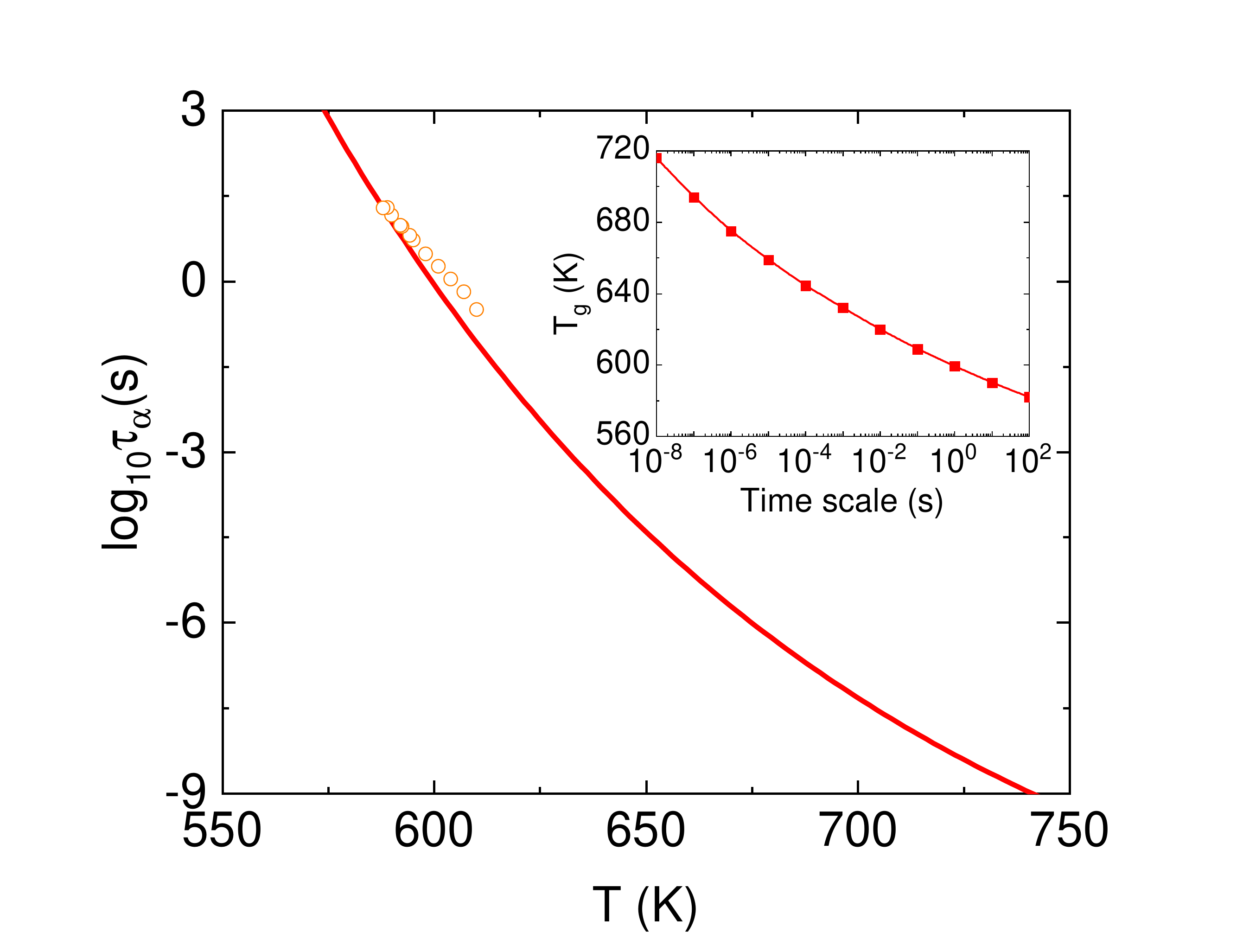}
\caption{(Color online) The temperature dependence of structural relaxation time of \ce{Pd_{40}Ni_{40}P_{20}}. Open points and a solid curve correspond to experimental data \cite{19} and ECNLE calculations, respectively. The inset shows the glass transition temperature as a function of vitrification time scales.}
\label{fig:1}
\end{figure}

Figure \ref{fig:1} shows the logarithm of theoretical and experimental structural relaxation time of bulk \ce{Pd_{40}Ni_{40}P_{20}} as a function of temperature. The experimental value of $T_g$ is 582 $K$ defined by $\tau_{\alpha,bulk}=100$$s$ and the theoretical $\tau_{\alpha,bulk}=100$ $s$ at $\Phi_g = 0.6585$. The set of parameters is used for our thermal mapping. A good quantitative agreement between theory and experiment is observed. Our ECNLE theory can also describe well with $\tau_\alpha(T)$ of simulations. As shown in Figure S1 in the Supporting Information, theoretical alpha relaxation time of bulk \ce{Cu_{50}Zr_{50}} and \ce{Cu_{46}Zr_{46}Al_{8}} are close to the simulation counterparts. Based on $\tau_{\alpha,bulk}(T)$, a variation of the theoretical bulk $T_g$ with vitrification criterion is calculated and shown in the inset of Fig. \ref{fig:1}. In simulation, the glass transition occurs when $\tau_\alpha= 10^3-10^6$ ps. %Although the value of $T_g$ in the simulation time scale is higher than that in the experimental observation, we can compare the theory results with experiment

\begin{figure*}[htp]
\includegraphics[width=18cm]{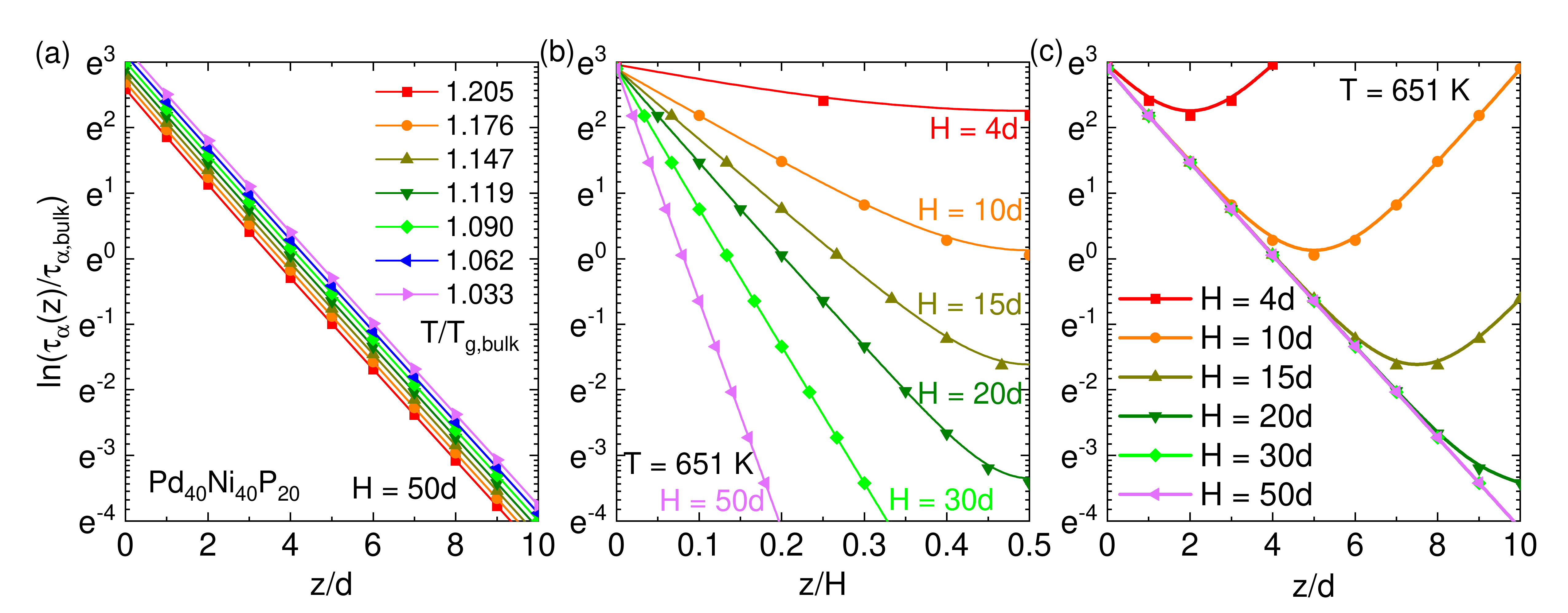}
\caption{(Color online) (a) Natural logarithm of normalized local structural relaxation time in a \ce{Pd_{40}Ni_{40}P_{20}} thin film having $H = 50d$ at different temperatures. $T_{g,bulk}$ is defined as $\tau_\alpha(T_{g,bulk}) = 100$ s. Natural logarithm of normalized mobility gradient at $T = 651 K$ and different film thickness (b) with and (c) without normalization by $H$.}
\label{fig:2}
\end{figure*}

Figure \ref{fig:2} shows the normalized mobility gradient of \ce{Pd_{40}Ni_{40}P_{20}} at different film thicknesses and temperatures. The dynamics is slowed down near the solid surface and is accelerated towards the film interior. One can see in Fig. \ref{fig:2}a that $\ln(\ln(\tau_\alpha(z)/\tau_{\alpha,bulk}))$ is linearly proportional to $z/d$. This form is known as the double exponential behavior and is found in a prior simulation working on metallic glass films \cite{14}. For sufficiently thick films, the dynamic free energy in Eq. (\ref{eq:4}) has the length scale or decay length, $\xi=d/\ln(2)\approx 1.44d$, which is independent of density or temperature. Its consequences leads to $\tau_\alpha(z)/\tau_{\alpha,bulk}$ decays in a weakly temperature-dependent manner but this result disagrees with Ref. \cite{14}. The disagreement suggests that atomic cooperativity or collective motion effects on glassy dynamics of metallic glasses can be very small but cannot be ignored. It still plays a role in metallic glasses. Comparisons between ECNLE calculations and simulations for \ce{Cu_{50}Zr_{50}} films in Ref. \cite{14} are shown in Figure S2 in the Supporting Information.

The gradient of normalized local relaxation time in a very thin film ($H = 4d$) is flattened due to strong interfacial interference between two solid surfaces as shown in Fig. \ref{fig:2}b and c. The double exponential behavior clearly occurs near the solid surface but the gradient is roughly flattened in the film center. This means $\ln(\tau_\alpha(z,H)/\tau_{\alpha,bulk})\approx$ constant at a given thickness as $z\rightarrow H/2$. This result completely agrees with simulations in Ref. \cite{14}. The flattening of dynamical gradients in the midfilm region has been predicted by both simulation and theory for polymer free-standing films \cite{24}. In thicker films, the interference becomes smaller and the mobility gradient approximately behaves as if an one-interface system. The deviation between dynamics near the solid surface and in the film interior is large enough to remove the flattening.

In metallic glasses, the density is very high and atomic structure is tightly packed. We can use the analysis of the ultra-local limit \cite{12}, which assumes a dominant role of large wavevector in calculations of $F_{dyn}(r,z,H)$, to obtain an analytical expression of $r_L(z,H)$. In this limit, $S(q)\approx 1$ and $C(q) = -4\pi d^3 g(d)\cfrac{\cos(qd)}{(qd)^2}$, with $g(d)$ being the contact number of a particle. Thus, we have
\begin{eqnarray}
\frac{r_L(z,H)}{r_{L,bulk}} = \frac{1}{1+\cfrac{\sqrt{2}-1}{2}\left(\cfrac{1}{2^{z/d}}+\cfrac{1}{2^{(H-z)/d}}\right)}.
\label{eq:6}
\end{eqnarray}

Then, the dynamic shear modulus is \cite{12}
\begin{eqnarray}
G(z,H) &=& \frac{9\Phi k_BT}{5\pi r_L(z,H)^2d}=G_{bulk}\left(\frac{r_{L,bulk}}{r_L(z,H)} \right)^2 \nonumber\\
&=& G_{bulk}\left[1+\frac{\sqrt{2}-1}{2}\left(\cfrac{1}{2^{z/d}}+\cfrac{1}{2^{(H-z)/d}}\right)\right]^2,
\label{eq:7}
\end{eqnarray}
where $G_{bulk}$ is the bulk dynamic shear modulus. The shear modulus can be used in the shoving model to provide another analytical interpretation for the double-exponential behavior of the mobility gradient. The alpha time is now given by
\begin{eqnarray}
\ln\left(\tau_\alpha(z,H)\right) = \ln\tau_{c} + \frac{G(z,H)V}{k_BT},
\label{eq:8}
\end{eqnarray}
where $\tau_{c}$ is a characteristic time and $V$ is the activation volume. For simplification, we suppose that the activation volume does not change compared to its bulk counterpart. From this, 
\begin{eqnarray}
\ln\left(\frac{\tau_\alpha(z,H)}{\tau_{\alpha,bulk}}\right) &=& \frac{G_{bulk}V}{k_BT}\left[\frac{G(z,H)}{G_{bulk}}-1\right] \nonumber\\ &=&  \ln\left(\frac{\tau_{\alpha,bulk}}{\tau_{c}}\right)\left[\frac{\sqrt{2}-1}{2}\left(\cfrac{1}{2^{z/d}}+\cfrac{1}{2^{(H-z)/d}}\right)\right]^2. \nonumber\\
\label{eq:9}
\end{eqnarray}
Clearly, $\ln(\tau_\alpha(z,H)/\tau_{\alpha,bulk}) \sim e^{-2z/\xi}+e^{-2(H-z)/\xi} \sim e^{-2z/\xi}$ when $H$ is large. This analysis suggests the derivation for a linear superposition of dynamical thick-film gradients empirically used in various works \cite{24,17}. 

\begin{figure*}[htp]
\includegraphics[width=8.5cm]{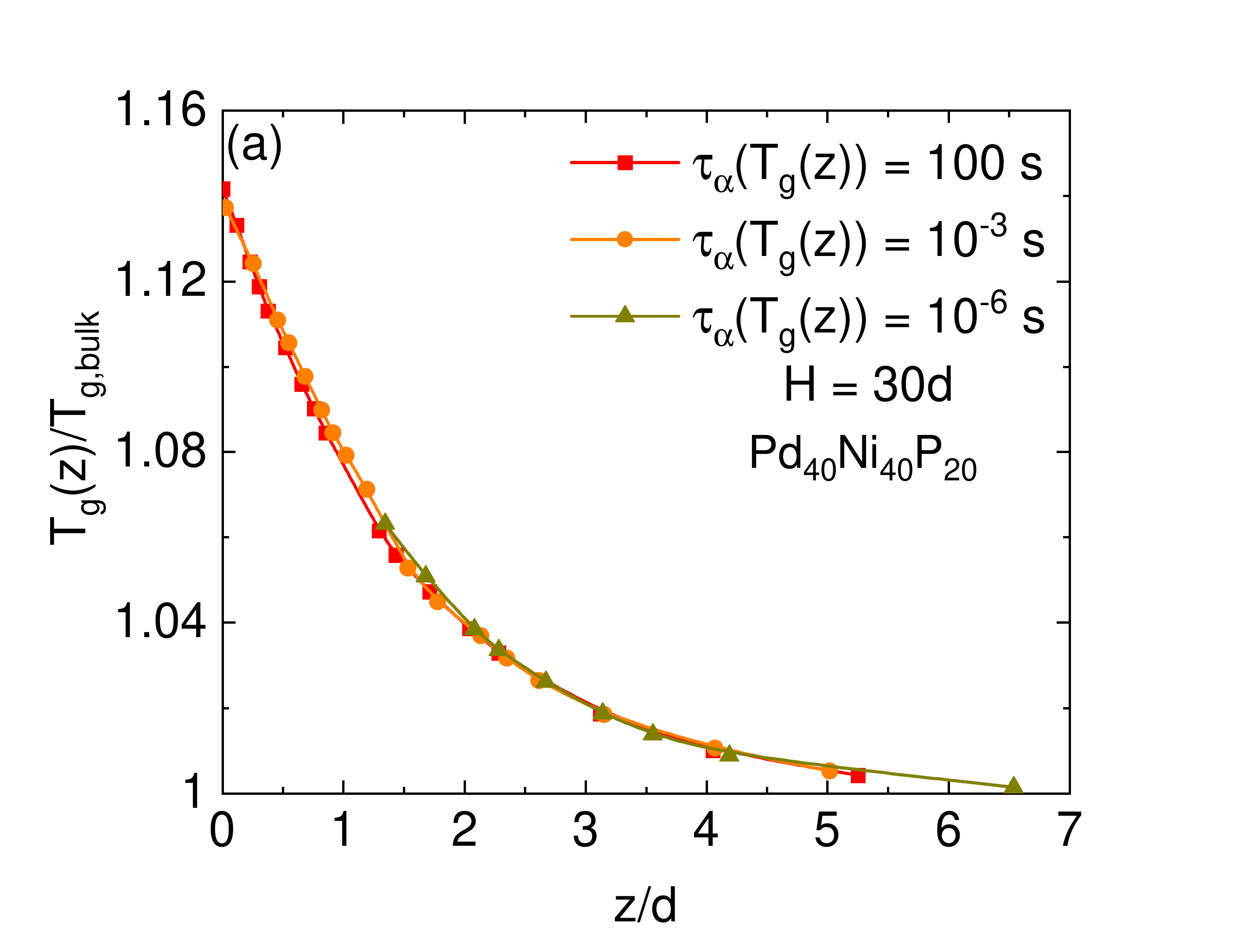}
\includegraphics[width=8.5cm]{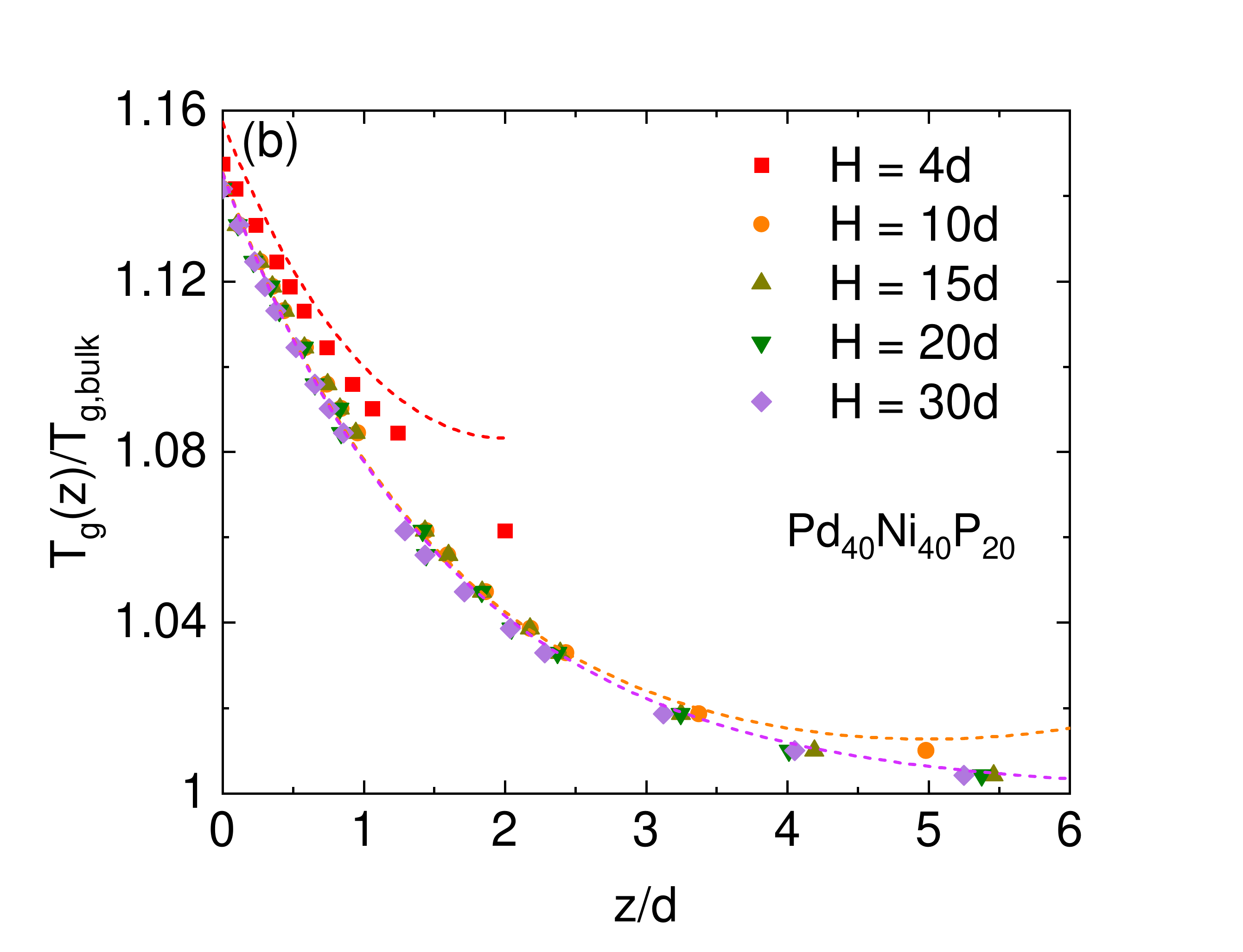}
\caption{(Color online) (a) The local glass transition temperature normalized by its bulk counterpart calculated using different vitrification time scales ranging from $10^6$ ps to 100s at $H = 30d$. (b) The local glass transition temperature normalized by its bulk counterpart (defined by the vitrification time scale of 100 s) in different film thicknesses. Solid data points and dashed curve corresponds to theoretical calculations and fitting curves.}
\label{fig:3}
\end{figure*}

Figure \ref{fig:3} shows the gradient of the normalized glass transition temperature at several film thicknesses and vitrification criteria. We calculate $T_{g}(z,H)/T_{g,bulk}$ as a function of $z/d$ at $H = 30d$ using $\tau_\alpha(T_g(z,H))=$ $10^{6}$ ps, 1 ms, and 100 s. This range of timescale covers from simulation to experiment. As seen in Fig. \ref{fig:3}a, the normalized $T_g$ gradients are almost perfectly overlapped. This finding reveals that although simulation is inaccessible to experimental observation or low temperature regimes, $T_{g}(z,H)/T_{g,bulk}$ predicted by simulation can describe well experimental data. Numerical results in Fig. \ref{fig:3}b shows that changing the film thickness slightly affects the normalized $T_g$ gradient when $H \geq 10d$. A significant deviation is found when the metallic glass film is very thin ($H = 4d$), where the spatial transfer of slowing down of dynamics from two surfaces is dynamically coupled. Our theoretical data can be fitted using the superposition approximation \cite{17}
\begin{eqnarray}
\frac{T_{g}(z,H)}{T_{g,bulk}}=1+A\left(e^{-z/\xi_L}+e^{-(H-z)/\xi_L}\right),
\label{eq:10}
\end{eqnarray}
where $A$ and $\xi_L$ are adjustable fit parameters. The first two terms $(1+Ae^{-z/\xi_L})$ are a fitting function of the normalized $T_g$ gradient of very thick films and the last term is responsible for effects of another surface on the glass transition. Fitting our data for a film of $H = 50d$, which can be considered as a thick film, with Eq. (\ref{eq:10}) gives $A =0.145$ and $\xi_L=1.6d$. As seen in Fig. \ref{fig:3}b, the superposition approximation provides a good description for metallic glass films having $H \geq 15d$, while it fails to predict $T_{g}(z,H)/T_{g,bulk}$ at $H = 4d$. This is because this treatment underestimates the interfacial coupling at solid surfaces of thin films and its consequences in the film interior.

\begin{figure*}[htp]
\includegraphics[width=8.5cm]{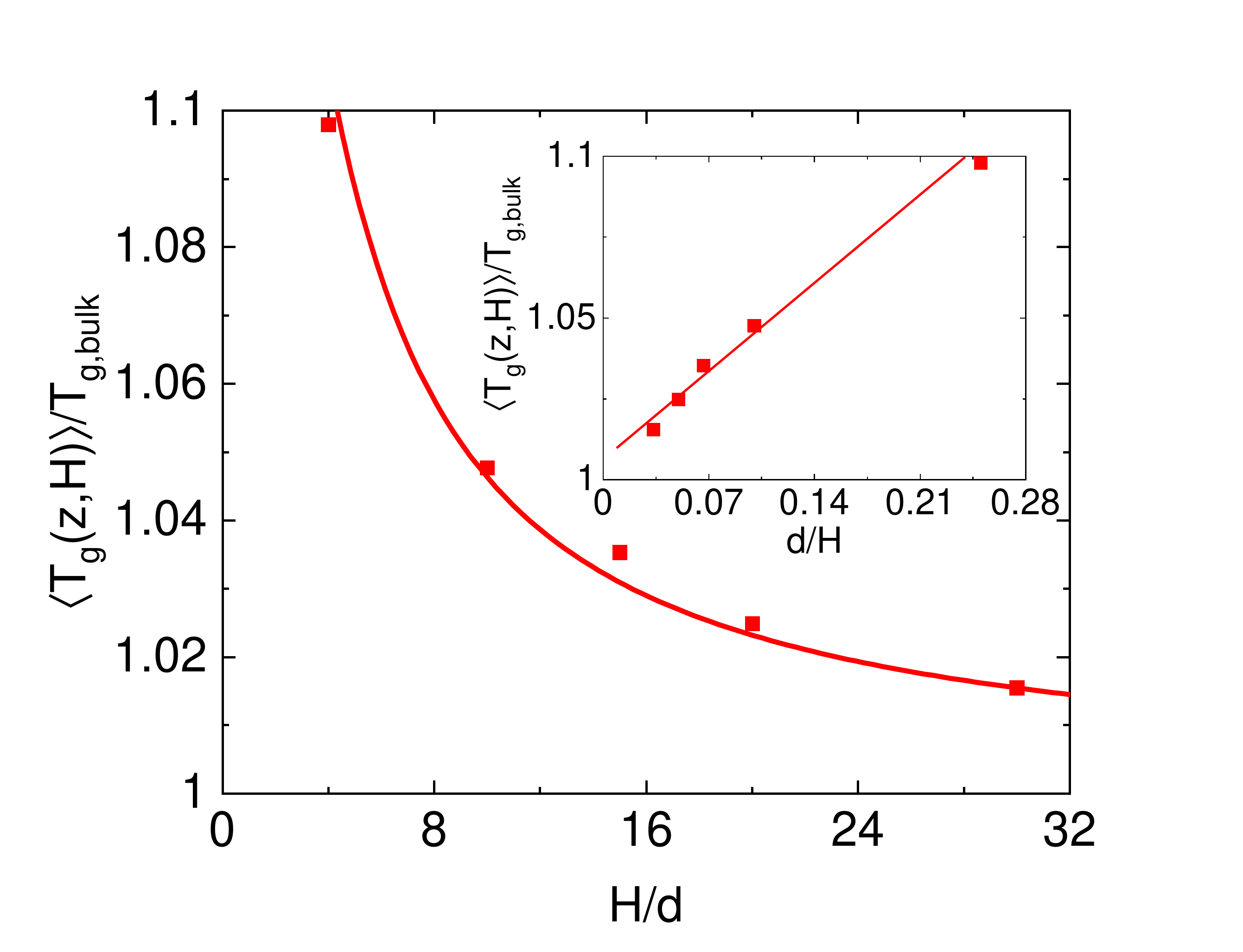}
\caption{(Color online) The film-averaged $T_g$ normalized by its bulk value as a function of $H/d$ for the 100s vitrification criterion. The solid curve is a fit using Eq. (\ref{eq:11}) with $A =0.145$ and $\xi_L=1.6d$. Inset: same results as in the mainframe plotted versus $d/H$. A straight line is a guide to the eye.}
\label{fig:4}
\end{figure*}

Based on Eq. (\ref{eq:10}), the film-averaged $T_g$ is
\begin{eqnarray}
\frac{\left<T_{g}(z,H)\right>}{T_{g,bulk}}=1+\frac{2A\xi_L}{H}\left(1-e^{-H/\xi_L}\right).
\label{eq:11} 
\end{eqnarray}
As can be seen in Fig. \ref{fig:4}, Eq. (\ref{eq:11}) quantitatively describes the thickness-dependent film-average $T_g$ of capped films predicted by the ECNLE theory. When $H \gg \xi_L$, Eq. (\ref{eq:11}) suggests $\left<T_{g}(z,H)\right>/T_{g,bulk}\sim 1/H$ and the inset of Fig. \ref{fig:4} numerically confirms this linear relationship. The linear form is not exact since there are deviations at very thin and thick films, but the description on intermediate length scales is relatively good. This behavior is also consistent with experiments in Ref. \cite{18} and completely inverse to the case of free-standing film \cite{15,16}. However, when $H \ll \xi_L$, $\left<T_{g}(z,H)\right>/T_{g,bulk}=1+2A$ is independent of the film thickness and it suggests that the mobility gradient is perfectly flattened.

\begin{figure}[htp]
\includegraphics[width=8.5cm]{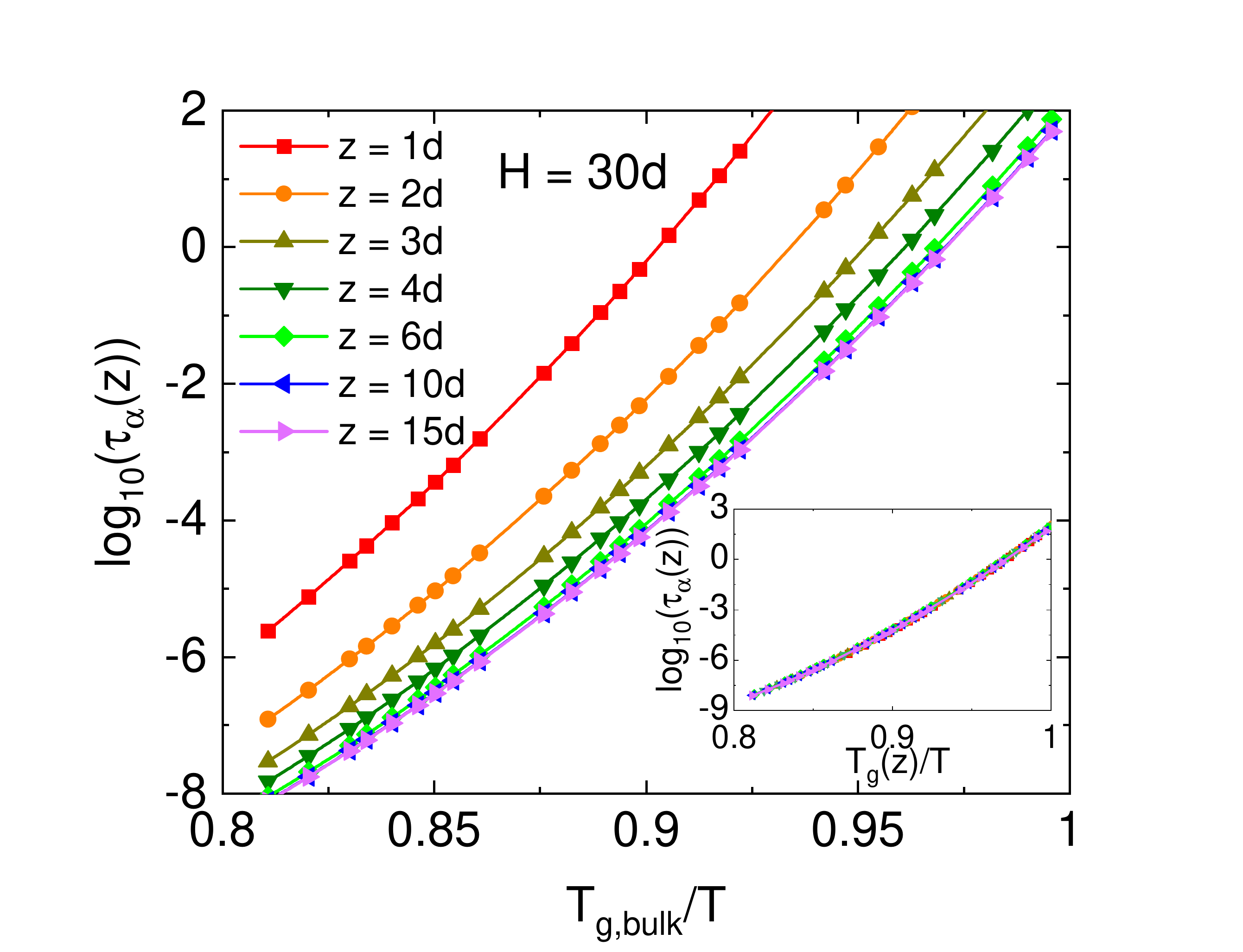}
\caption{(Color online) The temperature dependence of the structural relaxation time (secs) at various indicated distances from a solid surface of the capped \ce{Pd_{40}Ni_{40}P_{20}} film having $H = 30d$ and $T_{g,bulk}=582$ $K$. The inset shows the same data in the mainframe but as a function of $T_g(z)/T$, here $\tau_\alpha(T_g(z)) = 100$s.}
\label{fig:5}
\end{figure}

The mainframe of Fig. \ref{fig:5} shows the alpha time at certain $z$ values as a function of $T_{g,bulk}/T$. This is another way to see a significant growth of relaxation near the solid surface and interfacial effects on the caging constraints decay in the film middle. When $z \geq 10d$, $\tau_\alpha(z)$ slightly changes and seems to recover to its bulk. To understand how the dynamic fragility is spatially varied, we plot $\tau_\alpha(z)$ versus $T_g(z)/T$ in the inset of Fig. \ref{fig:4}. Remarkably, all data perfectly overlapped and this finding implies that there is no change in the local dynamic fragility compared to the bulk counterpart. Moreover, the gradient mobility can be calculated using the $T_g$ gradient and vice versa.
\newpage
\section{Conclusions}
In conclusion, we have developed the ECNLE theory to investigate the gradient of structural relaxation time and glass transition temperature in finite-size capped films of metallic glasses. The temperature dependence of $\tau_{\alpha,bulk}$ is in a good quantitative agreement with experiment. The presence of solid surfaces rigidifies the caging constrants at the interface and, thus, slows down the dynamics nearby. We find that the mobility and $T_g$ gradient normalized by their bulk values are well described using the superposition approximation. When the film is sufficiently thin, the flattening of the relaxation gradient occurs due to strong interference of interfacial effects. For thicker films, the dynamic coupling between two surfaces become minor and the double exponential behavior of relaxation reported by previous simulation \cite{14} as probing several layers near the interface is explained. However, our exponential decay length nearly remains unchanged with temperature and this result disagrees with Ref. \cite{14}. This is a consequence of ignoring contribution of collective motions to the glass transition. The ratio of $T_g(z)/T_{g,bulk}$ is unaffected by the vitrification criteria, which define for $T_g$ of simulation and experiments. Although the solid surfaces change both $\tau_\alpha(z)$ and $T_g(z)$, the Angell plot shows perfect overlapping among data of different layers in the film. This finding suggests a correlation between local relaxation and glass transition temperature.

\section*{Supporting Information}
ECNLE calculations for the alpha relaxation time of bulk and films of \ce{Cu_{50}Zr_{50}} and \ce{Cu_{46}Zr_{46}Al_{8}} at different temperatures are given and contrasted with their simulation counterparts.

\section*{Acknowledgement}
This research was funded by the Vietnam National Foundation for Science and Technology Development (NAFOSTED) under grant number 103.01-2019.318.

\section*{Supporting Information}
\begin{figure}[htp]
\includegraphics[width=8.5cm]{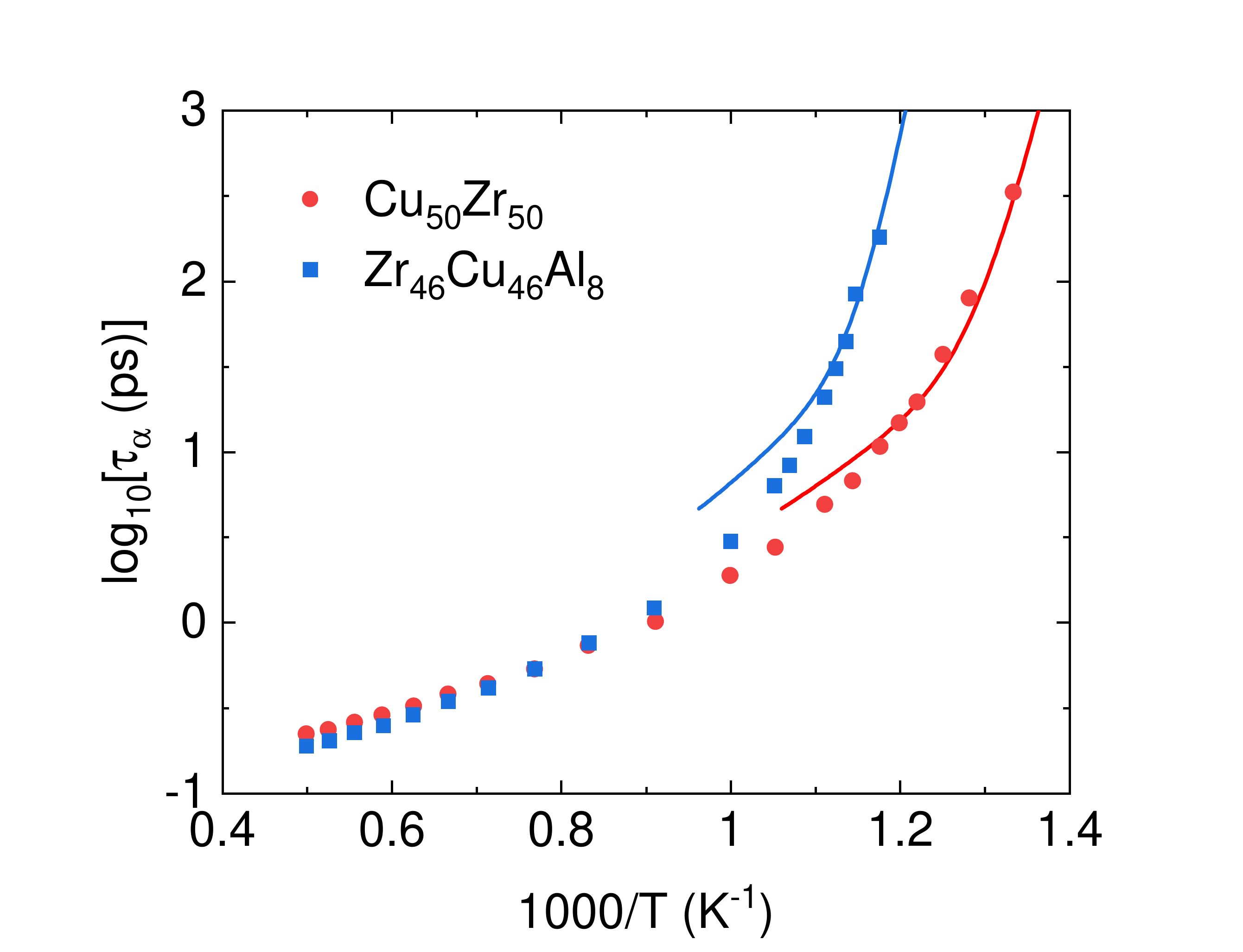}
\caption{\label{fig:SM1}(Color online) The temperature dependence of the bulk alpha time of \ce{Cu_{50}Zr_{50}} and \ce{Cu_{46}Zr_{46}Al_{8}}. Data points and solid curves correspond to simulation in Ref.\cite{14} and ECNLE calculations, respectively.}
\end{figure}

Figure \ref{fig:SM1} shows $\tau_\alpha(T)$ of the bulk \ce{Cu_{50}Zr_{50}} and \ce{Cu_{46}Zr_{46}Al_{8}} calculated using ECNLE theory and molecular dynamics (MD) simulations in Ref. \cite{14}. To convert from density to temperature, we still use the thermal mapping $T = T_{g,bulk} + (\Phi_g-\Phi)/\beta\Phi_0$ as described in the main text. However, we employ $T_{g,bulk} = 733.68$ $K$ and 829.2 $K$ for \ce{Cu_{50}Zr_{50}} and \ce{Cu_{46}Zr_{46}Al_{8}}, respectively, which is defined by $\tau_\alpha(T_{g,bulk})=1$ $ns$ or $10^3$ $ps$. Based on ECNLE calculations for bulk hard sphere systems, $\Phi_g$ is found to be 0.576 to have $\tau_\alpha(\Phi_g)=1$ $ns$. Theoretical results quantitatively agree well with simulations. 

\begin{figure}[htp]
\includegraphics[width=8.5cm]{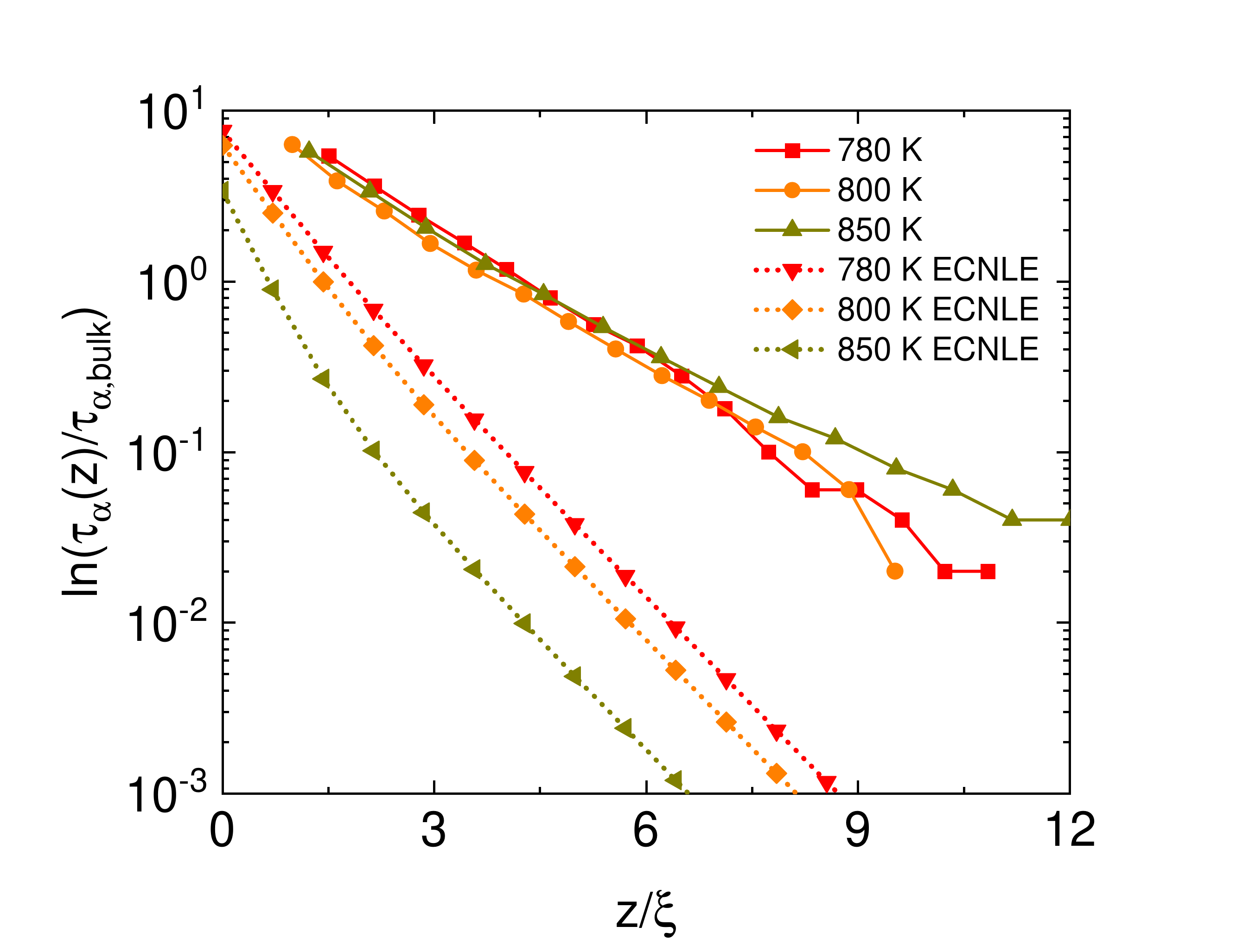}
\caption{\label{fig:SM2}(Color online) Natural logarithm of normalized local structural relaxation time in \ce{Cu_{50}Zr_{50}} thin films at $T = 780$, 800, and 850 $K$ as a function of $z/\xi$. Solid and dotted curves corresponds to simulation in Ref.\cite{14} and ECNLE results for film thickness of $H = 100d$, respectively.}
\end{figure}

Figure \ref{fig:SM2} shows $\ln(\tau_\alpha/\tau_{\alpha,bulk})$ of \ce{Cu_{50}Zr_{50}} thick films calculated using ECNLE theory and molecular dynamics (MD) simulations in Ref. \cite{14}. Both theory and simulation predict the double exponential form of spatial gradients of molecular mobility in the metallic film. In addition, when $z$ is scaled by the decay length, the slopes of theory and simulation curves reveal temperature insensitivity. However, simulation curves nearly collapse into a master curve while the theory curves are not overlapped. A main reason is the decay length of ECNLE calculations is temperature-dependent, meanwhile that of simulations is thermally varied \cite{14}. This is because we completely ignore effects of collective dynamics on the glass transition of metallic glasses. 

\end{document}